\documentstyle[prb,aps,multicol,epsf]{revtex}
\begin{document}
\draft
\title{Superexchange via Cluster States: \\ 
Calculations of Spin-Phonon Coupling Constants for CuGeO$_3$}
\author{S. Feldkemper, W. Weber}
\address{Institut f\"ur Physik, Universit\"at Dortmund, 44221 Dortmund,
Germany} 
\date{\today}
\maketitle
\begin{abstract}
Calculations for spin-phonon coupling constants in CuGeO$_3$ are
presented, applying fourth order superexchange perturbation theory to an
extended Hubbard model. In our treatment, the intermediate oxygen
ligand states are described by band-like cluster states, due to the
presence of significant O(p)-O(p) hopping. We also include the effect
of the Ge ions on the O ligand states.

We find a considerable q-dependence of the spin-phonon coupling. For
the ${\pi}$-modes involved in the spin-Peierls transition, our results
of the coupling constants are in fair agreement with the work of
Werner, Gros, and Braden. Yet some discrepancies remain, especially
concerning the coupling to the vibrations, which modulate the
O(2)-O(2)-Ge angle ${\delta}$. Our studies of the pressure dependence
of the magnetic coupling constants $J_1$ and $J_2$ suggest, that
relatively large non-linear effects are present, even at small
pressure values, probably due to the existence of a `soft',
van-der-Waals-type bond direction in CuGeO$_3$.
\end{abstract}	
\pacs{12.38.Bx, 31.15.Md, 63.20.Kr, 71.27.+a, 75.10.-b, 75.30.Et,
75.50.-y, 75.50.Ee} 
\begin{multicols}{2}
\narrowtext
%
%
\section{Introduction} 
%
%
In recent years, the magnitudes and origin of the magnetic coupling
constants of the quasi-one-dimensional quantum spin system CuGeO$_3$
have been subject of a variety of papers
\cite{jest,riera95,khomskii96,fab98,brad98}. In particular, the
anti-ferromagnetic first nearest neighbor coupling $J_1$ is unusually
small - believed to arise from superexchange hopping paths forming
Cu-O(2)-Cu angles ${\eta}$ close to $90^{\circ}$, which is the
Goodenough-Kanamori-Anderson \cite{gka} (GKA) criterion for vanishing
anti-ferromagnetic superexchange. 

Some authors have suggested that the hopping paths are strongly
modified by the influence of the Ge atoms on the O(2) ions, because of
the strong covalent ionic Ge-O(2) bonds. \cite{khomskii96} Other
authors \cite{brad98} have obtained very strong variations of $J_1$
with the angle 
${\eta}$; i.e., have predicted very large spin-phonon coupling
constants in this system, in contradiction to estimates based on the
data of the spin-Peierls transition. \cite{riera95.2}

Very recently, Werner, Gros, and Braden (WGB) have presented a
detailed evaluation of experimental data, based on a variety of
theoretical estimates, to derive the spin coupling to the four phonon
modes involved in the spin Peierls transition. \cite{WG99} In
particular, they find a strong coupling to variations of the angle
${\eta}$, and a relatively large coupling to the Ge displacements.

In the present paper we use superexchange perturbation theory to evaluate the
spin coupling constants between the magnetic ions and their modulations
with atomic displacements. Our method is an extension of earlier
perturbative treatments. \cite{FeWe98} In particular, we include
band-like oxygen ligand states, because there is significant direct
O(2)-O(2) hopping in the chain.  For our calculations, we start from a
Hubbard-type Hamiltonian with five d-orbitals ${\alpha},{\beta}$ per
Cu site and p-orbitals at the O ligand sites. The full number of d
orbitals at the Cu sites is considered in order to allow both for
symmetry induced orbital mixing and for the possibility of
ferromagnetic exchange coupling involving excited Cu d levels.
On the oxygen sites, we include the splitting and mixing of the
oxygen p-levels caused by the Coulomb fields of the adjacent ions, in
particular by Ge.

The rest of the paper is organized as follows. In section II 
we discuss details of our model, in section III we first discuss
our choices of model parameters and then
present the results for the spin-phonon coupling constants and for the
pressure dependence of $J$. Concluding remarks are given at the end.
%
%
\section{Details of the model} 
\begin{figure}[p]
\begin{center}
\ \epsfxsize= 6 cm \epsffile{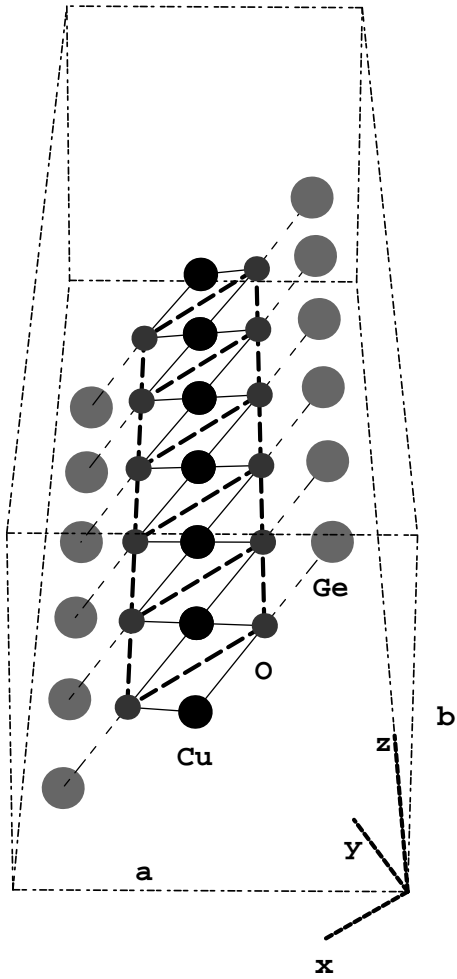}
\end{center}
\narrowtext
\caption{CuO$_2$ spin chains (along $z$ direction) with neighboring Ge
atoms. $x$ direction lies in the CuO$_2$ plane, $y$ perpendicular to
the plane. Also shown are the crystallographic axes $a$, $b$, and
$c=z$. Hopping paths are denoted by the heavy, broken line for O-O
hopping and the light, full line for Cu-O hopping. In addition, the
broken line between O and Ge indicates the ligand field of the Ge 
ions. } 
\label{cugeo_ladder} 
\end{figure}

%
%
The Cu atoms form a linear chain along the CuO$_2$ ribbons (see
Fig.~\ref{cugeo_ladder}), where the oxygen atoms in the CuO$_2$ plane are
labeled O(2), the apex oxygen atoms O(1).
    
We use a multi-band Hubbard-like Hamiltonian, where all electron-electron
interaction takes place at the Cu-site, for details see
Ref.~\onlinecite{FeWe98}. We take five d-orbitals
per Cu atom (sites $i$, orbitals ${\alpha}$) and 3 p-orbitals (sites
$l$, orbitals ${\tau}$) per oxygen and use the hole picture. We 
allow p-d hopping $t_{il}^{{\alpha}{\tau}} $ between Cu ions and next
nearest oxygen atoms. Also, oxygen-oxygen hopping
$t_{ll^{\prime}}^{{\tau}{\tau}^{\prime}}$ up to third nearest neighbors
is included. The Hamiltonian reads
\begin{eqnarray}
\label{hubbard}
H &=&  \sum_{\stackrel{i}{{\alpha},{\sigma}}}
               {\epsilon}^{\alpha}_i \hat{n}_{{\alpha}{\sigma}}(i) +
       \sum_{l,{\sigma}} {\epsilon}_l^{\tau} \hat{n}_{{\tau}{\sigma}}(l)
\nonumber \\ 
  &+&  \frac{1}{2} \sum_{\stackrel{l}{{\tau},{\tau}^{\prime},{\sigma}}}
       V_l^{{\tau}{\tau}^{\prime}}
   c^{\dagger}_{{\tau}{\sigma}}(l)c_{{\tau}^{\prime}{\sigma}}(l) \nonumber \\ 
  &+&  \sum_{\stackrel{i,l}{{\alpha},{\tau},{\sigma}}}
   [t_{il}^{{\alpha}{\tau}} c^{\dagger}_{{\alpha}{\sigma}}(i)
          c_{{\tau}{\sigma}}(l) + h.c.]   \nonumber  \\
  &+&  \sum_{\stackrel{l,l^{\prime}}{{\tau},{\tau}^{\prime},{\sigma}}}
   [t_{ll^{\prime}}^{{\tau}{\tau}^{\prime}}
   c^{\dagger}_{{\tau}{\sigma}}(l) c_{{\tau}^{\prime}{\sigma}}(l^{\prime}) +
              h.c.]          \nonumber  \\
  &+&  \frac{1}{2} 
       \sum_{\stackrel{\stackrel{i}{{\alpha},{\beta}}}
            {{\sigma},{\sigma}^{\prime}}}   
         U_{{\alpha}{\beta}}\
         c^{\dagger}_{{\alpha}{\sigma}}(i)
         c^{\dagger}_{{\beta}{\sigma}^{\prime}}(i)
         c_{{\beta}{\sigma}^{\prime}}(i)c_{{\alpha}{\sigma}}(i) \nonumber\\
  &-&  \frac{1}{2}       
       \sum_{\stackrel{\stackrel{i}{{\alpha}\neq{\beta}}}
               {{\sigma},{\sigma}^{\prime}}}
         {I}_{{\alpha}{\beta}}\
         c^{\dagger}_{{\alpha}{\sigma}}(i)
         c^{\dagger}_{{\beta}{\sigma}^{\prime}}(i)        
         c_{{\alpha}{\sigma}^{\prime}}(i)c_{{\beta}{\sigma}}(i).
\end{eqnarray}
${\epsilon}_i^{\alpha}$ and ${\alpha}$ are the respective on-site
energies of d and p orbitals. $U_{{\alpha}{\beta}}$ and
${I}_{{\alpha}{\beta}}$ are Coulomb and  exchange interactions at the
magnetic sites, respectively. In the atomic limit, the choice
${I}_{{\alpha}{\beta}} < 0$ leads to the first Hund's rule (maximum
spin), the choice $U_{{\alpha}{\alpha}} > U_{{\alpha}{\beta}}$ would
lead to the equivalent of the second Hund's rule (maximum orbital
momentum). Electron-electron interactions at the ligand ions and between
different sites are neglected. We will also use the approximation
$U_{{\alpha}{\alpha}} = U_{{\alpha}{\beta}} = U$ and
${I}_{{\alpha}{\beta}} =J_H$. We note that although this Hamiltonian
does not yield the correct atomic multiplet spectrum, it is
sufficiently general for our perturbative approach. In particular, it
incorporates the possibility for a competing ferromagnetic coupling of
spins.  

The main impact of the Ge
atoms on the electronic structure is to cause a significant crystal
field splitting of the oxygen p atomic levels, \cite{khomskii96} this
effect is approximated by a crystal field $
V_l^{{\tau}{\tau}^{\prime}}$ (including non-diagonal terms). Admixture of Ge
4s, 4p hole states to the O p hole states appears to be rather small
\cite{matt94} and thus is neglected here.  

Changes of hopping integrals $(ab{\gamma})$ due to variations of
inter-atomic distances away from the equilibrium
separation are considered using relations like $(ab{\gamma}) \sim
(ab{\gamma})_0 e^{{\beta}{(r_0-r)}}$, where $r_0$ is the respective
equilibrium distance.

The main difference to earlier perturbation theory calculations of
spin coupling constants lies in the handling of the ligand
ions. Usually, the ligands are treated in the atomic limit; i.e., only
those hopping processes are considered, which occur along the spin
exchange paths. Here we extend the method and include band-like ligand
states. This extension is justified by the observation that O-O
hopping integrals of considerable magnitude exist in the CuO$_2$ ribbons,
\cite{oxhop} leading to O(2) p-band widths of several eV. As we
investigate the spin coupling for finite size CuO$_2$ clusters, we
actually deal with a finite number of O(2) cluster states instead of
the complete set of band states.

Our calculations are carried out in a sequence of three steps. First,
the oxygen cluster states are determined. As mentioned above, only O(2)
p-orbitals are considered in the one-particle model. Input are
O(2)-O(2) hopping integrals, bare O(2) p-orbital energies and the
crystal field caused by the adjacent ions. The cluster states are then
obtained by diagonalizing the cluster Hamiltonian, using up to 16 O(2)
atoms and periodic boundary conditions. In a second
step,environmentally adapted Cu one-particle levels are
determined. This calculation is
carried out by coupling the atomic Cu levels (which are fivefold
degenerate) of a single Cu atom to the O(2) cluster and also to the
apical O(1) atoms. Here, the additional input are the p-d level separation and
the values of $pd{\sigma}$ and $pd{\pi}$ hopping integrals. The
coupling is calculated in second order perturbation theory and
leads to an effective on-site d-d hopping matrix, which is then
diagonalized. We note that a rather similar d-level ordering (with
respect to both d character and separation) can be obtained using
Coulomb-type ligand fields. In the third step, the adapted Cu d-levels
are then used to calculate the spin coupling constants by usual
fourth order superexchange perturbation theory, yet the Cu spins are
transferred via the O cluster states instead of atomic ligand
states. As the cluster (or band) states are delocalized, one might
expect that the interaction between the magnetic sites decays rather
slowly with distance.

This procedure yields all magnetic coupling constants from the same
order of perturbation theory, irrespective of the range of the
coupling. As we deal with delocalized ligand states, instead of atomic
ligand states, hopping processes are included up to infinite order, in
principle. Yet, this approach remains limited to the interaction of
pairs of magnetic sites. It should finally be noted that each hopping
path yields a certain amplitude, and a bilinear expression of
such amplitudes leads to the value of the respective coupling
constant. \cite{FeWe98} The
hopping paths may be divided into two classes, one where the
interacting spins meet at a magnetic site (the conventional picture of
superexchange), and the other where they meet in a cluster state. The
latter processes dominate, because of the presence of the many
cluster states. Interference effects play a very important role,
leading to some important consequences:
\begin{itemize}
\item{They limit the range of coupling to essentially first and second
nearest neighbors.}
\item{The ${\pi}/2$ destructive interference of the GKA rule is
removed (actually shifted in angle). Thereby the crystal fields at
the ligand sites, produced by the neighbor atoms (and modulated by
their motion) play an important role.\cite{khomskii96}}
\item{Electron-electron interaction at the ligand sites will also
modify the interference; however, there does not appear a tractable
way to include it into our perturbative approach.}
\end{itemize}  

Changes of atomic positions can be included easily in our
calculations, this allows to determine the variations of the nearest
and second nearest neighbor magnetic coupling constants ($J_1$ and
$J_2$) with atomic displacements. 
%
%
\section{Calculation of Spin-Phonon Coupling Constants} 
%
%
\subsection{Model Parameters}  

We have performed the calculations using cluster sizes of twelve and
sixteen oxygen ions and have used three different sets of parameters
(A, B, C). The values of the parameters lie in the range of values
discussed in the literature, e.g., for the hopping integrals we used
values similar to those given by Mattheiss. \cite{matt94} As our main
goal is to investigate the reasonable range of spin-phonon coupling
constants, we have chosen the model parameters to yield the standard
values of $J_1 \sim 160$ K and $J_2/J_1 = {\alpha} \sim
0.3$. \cite{jest,riera95,fab98,brad98} The model parameters are given in
Tab.~\ref{cugeoparsets}. Set A describes a model without coupling of
the Cu d levels to O(1). Sets B and C include the Cu d level coupling
to the apex oxygens. For set C smaller O-O hopping parameters than in
A and B have been used. The slight difference between O(2)-O(2)
hoppings along $x$ and $z$ is incorporated by the ${\beta}$
parameters; i.e.~by the distance dependence of $pp{\sigma}$ and
$pp{\pi}$. We consider sets A and B as typical for CuGeO$_3$, while set
C with its much smaller $pp{\sigma}$ value may indicate the range of
variations possible with still physically rather meaningful hopping
parameters. 

The crystal field acting on the O(2) p levels has been calculated
using a point charge model for a shell of seven neighbor ions. These
are O, Ge, Cu ions, where we assumed nominal ionic charges of $-2$,
$+4$, $+2$, respectively and a dielectric constant ${\varepsilon}=
6$. For the calculation of the crystal field integrals, hydrogen-like
2p wave functions of the O(2) sites were used. At equilibrium
separation, the monopole contribution (l=0) to 
the crystal field matrix was removed (as this term is incorporated in
the value of the atomic energy ${\epsilon}_p$). The resulting crystal
field splitting of the O(2) p levels is given in Tab.~\ref{penerg}. We
note that for atomic displacements, e.g., for the motion of the Ge
ions, the monopole terms are included in the calculation of the
spin-phonon coupling.  

Tab.~\ref{denerg} shows the Cu $d^9$ hole energies for the three sets of
parameters A, B, and C for 16 oxygen ions in the cluster.
In all cases, including the spin-phonon calculations, the Cu $d^9$
ground state is predominantly of $xz$ symmetry, with very little
admixture of the other d orbitals. The $d_{xz}$ orbital is always
lowered considerably, caused by the strong $pd{\sigma}$ hopping
between the $d_{xz}$ hole orbital 
and the oxygen-p cluster states. All other $d^9$ levels are much less
affected. Since for set C the p-p hopping is considerably smaller than
for sets A and B, the band width of the oxygen-p cluster states is
reduced and, as a consequence, also the d level splitting.
Experimental information on the positions of the excited Cu $d^9$
levels in CuGeO$_3$ is not known to us, a range between 1.5 eV and 3
eV appears to be reasonable. Our calculations of the magnetic coupling
constants also include the spin hopping paths involving excited Cu
$d^9$ levels. However, these contributions turn out to be negligible,
both for ferromagnetic and anti-ferromagnetic exchange paths. As a
consequence, the energy positions of the excited $d^9$ levels actually
turn out to be irrelevant for the values of the magnetic coupling constants.
\subsection{Spin-Phonon Coupling}  

The coupling constants in the equilibrium structure obtained with the
three parameter sets A, B, and C are 
displayed in Tab.~\ref{js}. For all sets, the
results for $J_1$ and $J_2$ are very similar. Typically, the
difference between 12 and 16 site clusters is negligible. We have also
determined values for $J_3$ and $J_4$, here we find $J_3/J_2 \sim
0.02$ and $J_4/J_3 \sim 0.1$. Therefore, we find {\it only} $J_2/J_1$
to be unusually large. If we further enhance the O(2) p band width by
using larger p-p hopping integrals, for example, we find larger ratios
$J_3/J_1$ and  $J_4/J_1$, i.e. longer range coupling constants.

For comparison with the work of WGB we have calculated the spin-phonon
coupling constants of the phonons involved in the spin-Peierls
distortion. For this purpose we have determined the change of $J_1$
with respect to 
\begin{itemize}
\item[(i)]{the motion of Cu along the crystallographic $c$ direction
($c =z$, see Fig.~\ref{cugeo_ladder}),}
\item[(ii)]{the motion of Ge along $b$,}
\item[(iii)]{the motion of O(2) along $a$,}
\item[(iv)]{the O(2) motion along $b$.}
\end{itemize}
Note that for (iii) and (iv) the two O(2) atoms bridging the Cu
chains move opposite to each other. Also note that along the chain
direction equivalent ions are displaced in alternating fashion
(${\pi}$-mode). Our values given in Tab.~\ref{coups} include a factor
$1/2$, for comparison with WGB, who used a different normalization.

A linear transformation leads to the angular coupling constants
$g_{\eta}$ (to the change of the Cu-O(2)-Cu angle ${\eta}$) and
$g_{\delta}$ (to the change of the O(2)-O(2)-Ge angle ${\delta}$) and
to the coupling to bond length variations $g_{Cu-O(2)}$ and
$g_{Ge-O(2)}$ (see Tab.~\ref{coupsphys}). The coupling to the four
Peierls-active phonons of $T_2^+$ symmetry \cite{brad98} can be
obtained by another linear transformation, using eigenvectors of these
modes as given by WGB (see Tab.~\ref{coupssp}).   

As we have used finite atomic displacements to determine the coupling
constants of Tab.~\ref{coups}, we have checked the accuracy of the
linear transformation by directly inserting the angular and bond
length variations to our program code. These results, also given in
Tab.~\ref{coupsphys}, agree very well with the others.

Tab.~\ref{coups} indicates that three of the four coupling constants
are of similar magnitude. These are $g_{Cu}^c$, $g_{O(2)}^a$, and
$g_{O(2)}^b$. All three quantities depend on the change of both the
Cu-O(2) bond length and on the O(2)-Cu-O(2) angle ${\eta}$, while
$g_{Ge}^b$ is mainly determined by the Ge-O(2) bond length
variation. As evident from Tab.~\ref{coupsphys}, $g_{\eta}$ is much
larger in magnitude than $g_{\delta}$ and is one of the dominant
coupling constants in this system. This observation also transforms to the
result of the spin-phonon coupling constants of
Tab.~\ref{coupssp}. Here, $g_1$ represents the coupling to the
lowest-lying phonon mode, which mainly involves variations of the
angle ${\delta}$. The coupling $g_2$ is particularly strong, to a large
extend this mode incorporates ${\eta}$ angle distortions.

In Tab.~\ref{coupgz}, some spin-phonon coupling constants for $q_z=0$
displacements are given (now also including values for $J_2$, which
are not accessible for ${\pi}$-modes). We observe a significant $q$
dependence of the $J_1$ coupling constant of order 15-30 \%.

We note that the WGB results differ considerably from our
calculations. The differences may be best discussed looking at
Tab.~\ref{coupsphys}. While the $g_{\eta}$ values agree
satisfactorily, the $g_{\delta}$ values are larger by an order of
magnitude in the WGB paper. The two other coupling constants
$g_{Cu-O(2)}$ and $g_{Ge-O(2)}$ are comparable in magnitude, yet
exhibit a different sign. 

We have also determined values for the spin-Peierls temperature
$T_{SP}$ using the same approach as employed by WGB. All our sets of
coupling constants $g_i$ yield much smaller values of $T_{SP}$. 

We presume that the difference in the results for $g_{\delta}$ is the
key to the discrepancies. As the ${\delta}$-mode is the lowest-lying
mode of the four Peierls active phonons, yet strongly involves
bond-length and bond-angle ${\eta}$ conserving displacements of the
light oxygen ions, we suspect strong anharmonic contributions
especially to $g_{\delta}$. These contributions may show up as
pseudo-harmonic in the WGB evaluation - thereby, via eigenvector sum
rules of the harmonic problem, also effecting the other coupling
constants. We note that our pressure results below also indicate the
presence of non-linear effects, even at small pressures.  

An earlier calculation of $g_{\eta}$ using exact diagonalization
techniques performed by Braden et al. \cite{brad98}
has produced results for the spin-phonon coupling which are
considerably larger than our results. The calculation was performed
using a cluster of two Cu ions, two oxygen ions, and two Ge ions. 

The biggest difference between our model and that of
Ref.~\onlinecite{brad98} probably is their use of an on-site
interaction $U_p$ at the ligand sites. This interaction affects all
spin hopping amplitudes (in our perturbative picture), it therefore
reduces the interference effects leading to the GKA
minimum. Additionally, there may exist a problem concerning the signs
of the p-p hopping integrals. Using the model parameters of
Ref.~\onlinecite{brad98}, we find reasonable agreement to our results
only, if we change the signs of the p-p hopping integrals. (Note that
our choice of sign is consistent - in the hole picture - with the
assumption that the cluster states of predominantly ${\sigma}$
bonding type are lowest in energy.) 

We have also determined the value of the exchange constant splitting
${\delta}_{SP} = |J_1^a -J_1^b|/(J_1^a + J_1^b)$, 
i.e.~the difference between the nearest neighbor couplings in two
adjacent cells in the dimerized phase. We find ${\delta}_{SP} =
0.012$. This is in good agreement with the results of Riera and Dobry,
\cite{riera95} who report ${\delta}_{SP} = 0.014$. Note that
${\delta}_{SP} \sim 
0.012$ (together with $J_2/J_1 = {\alpha} = 0.35$) is necessary to
reproduce the spin gap ${\Delta} = 2.1$ meV. \cite{buech98} 

\subsection{Pressure dependence of $J_1$ and $J_2$}  

The distortions of the seven crystallographic parameters under pressure have
been studied by various groups. \cite{brad99,brau97,adams} While the
pressure data on $a$, $b$, and $c$ are quite well established, there
are somewhat larger discrepancies in the results for the four internal
lattice parameters such as the value $x$ for O(2). These data also
show large scattering, especially at low pressures. Additionally, we
note a relatively large sample dependence of the $p=0$
crystallographic data.

\begin{figure}[p]
\begin{center}
\ \epsfxsize= 8 cm \epsffile{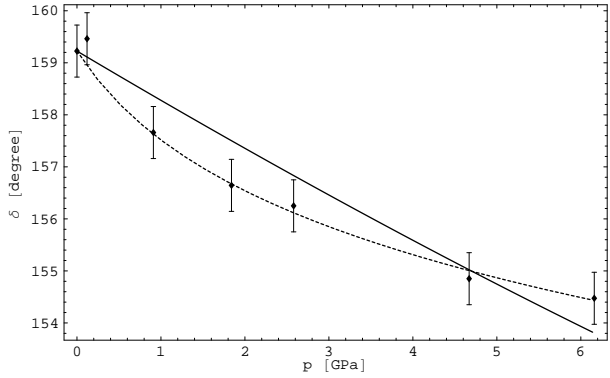}
\end{center}
\narrowtext
\caption{Dependence of the O(2)-O(2)-Ge angle ${\delta}$ on pressure. The
symbols show the experimental values, the full (dotted) line shows
${\delta}(p)$ calculated from a linear (Murnaghan-like) fit to the
experimental crystal parameters. \protect\cite{brad99}}
\label{delta}
\end{figure}

We have used two approaches to incorporate the experimental data,
especially those of Braden et al., \cite{brad99} into our calculation:
\begin{itemize}
\item[(i)]{we use a linear interpolation}
\item[(ii)]{we allow non-linear dependence of the experimental data
on pressure, using Murnaghan-type relations, \cite{murn44} as
previously proposed by Br\"auninger et al.~for
CuGeO$_3$. \cite{brau97} This is motivated by the existence of the
`soft' $b$ direction, where a van-der-Waals coupling dominates, see
Fig.~\ref{delta}.}
\end{itemize}
Our results depend strongly on the kind of interpolation of the experimental
pressure data. The $p=0$ gradients differ considerably (see
Tab.~\ref{prcoup}), especially affected are lattice constants involving
the `soft' $b$ direction. The resulting logarithmic derivatives of
$J_1$ and $J_2$ with 
respect to pressure deviate by almost an order of magnitude, see
Fig.~\ref{jp}. Yet, even using the linear approach, the variation of
$J_1$ with pressure is rather non-linear. 

For $\frac{{\partial} {\ln} J_1}{{\partial} p}$
we find -1.6 \%/GPa for the linear interpolation and -10.7 \%/GPa for the
non-linear one. In Ref.~\onlinecite{Jpexp}, an experimental value of
-8 \%/GPa has been reported. For $\frac{{\partial} {\ln}
J_2}{{\partial} p}$ we obtain 1.1 \%/GPa (linear) and 9.6 \%/GPa
(non-linear). In a theoretical study using density matrix
renormalization group techniques, Raupach et al.~obtain $\frac{{\partial} {\ln}
J_2}{{\partial} p} \sim 15 $ \%/GPa (under the assumption
$\frac{{\partial} {\ln} J_1}{{\partial} p} = $ -8 \%/GPa and ${\alpha}
\sim 0.3 $). \cite{raup99} WGB report a value of $\frac{{\partial} {\ln}
J_1}{{\partial} p} \sim -6 $ \%/GPa, however they extrapolate all
coupling constants to the homogeneous deformation of the lattice from
the four ${\pi}$-mode spin-phonon coupling constants, neglecting the
q-dependence of the coupling. 
 
In summary, our pressure results are based on an incomplete
experimental data base. Nevertheless, there are indications that
there exists a considerable non-linearity in the pressure dependence,
even at small pressures, probably due to the `soft' b-direction of the
crystal.   
%
%
\section{Conclusion} 
In this paper we present results of calculations for spin coupling
constants in CuGeO$_3$. In an extension of earlier work we have
applied fourth order superexchange perturbation theory to a
Hubbard-type model with 
five d orbitals at the Cu sites and three p orbitals at the O ligand
sites. Because of the significant direct p-p hopping, delocalized O(2)
cluster states have been used. In this way, hopping processes up to
infinite order have been included, equivalent, in principle, to the
RKKY treatment of the interaction of spin pairs in metals.
In our treatment we also incorporate the effects of the Ge sites, which
mainly cause splittings of the O(2) p levels and modulate them with
atomic displacements.

Our main results are:
\begin{enumerate}
\item{The largest contributions to the spin-exchange coupling constants 
arises from those hopping paths, where doubly occupied cluster states
are involved. Doubly occupied Cu states are less
important. Further, only processes involving the Cu d ground-state
level contribute significantly, all processes involving excited Cu d levels
can be ignored.}  
\item{The range of spin coupling constants is limited to $J_1$ and
$J_2$. Any longer distance coupling can be ignored for the range of
parameters we used for our calculations. Increasing the oxygen band
width (by increasing the p-p hopping
integrals) leads to a longer range of the magnetic coupling.}
\item{The spin-phonon coupling constants show considerable
q-dependence. The results for the ${\pi}$-modes are in satisfactory agreement
with the values presented by WGB. There are, however, also
significant differences, in particular, concerning the magnitude of
the coupling to the angle ${\delta}$ (O(2)-O(2)-Ge). We have
investigated a certain range of O(2)-O(2) hopping integrals (in limits
which we believe are physically meaningful), yet we could not find
significantly better agreement with WGB results.}
\item{Calculations of the pressure dependence of $J_1$ and $J_2$ yield
very different results depending on the interpolation
(either linear or non-linear) of the crystallographic data. The
results using a non-linear interpolation agree much better with
experimental data and other theoretical estimates of the pressure
dependence of $J_1$ and $J_2$.}
\end{enumerate}

Although the crystallographic pressure data exhibit significant
scattering, in particular for the internal lattice parameters, and
further appear to be sample dependent to some extend, we think that
the non-linear pressure dependence may, indeed, be a real effect. The
CuGeO$_3$ lattice structure exhibits a `soft' direction along the $b$
axis with van-der-Waals-type bonding between the sheets containing the
CuO$_2$ chains. Crystals with similarly anisotropic bonding, such as
the spiral structures of Se or Te also show significant non-linearity
in the pressure data. \cite{richt76} It would be highly desirable to
have available improved crystallographic pressure data.

Non-linear, i.e.~anharmonic behavior of certain phonon modes may also
cause some of the discrepancies between our results on spin phonon
coupling and those of WGB. A candidate for large anharmonicities is
the lowest lying mode of the four phonon modes involved in the
spin-Peierls transition. This mode mainly consists of O(2)
displacements, which modulate the O(2)-O(2)-Ge angle ${\delta}$. 

Finally, we should remark that some of the assumptions underlying our
studies may be questionable. This does not so much concern the values
of the model parameters - we think, we have studied a physically
meaningful range. It concerns more the assumption that on-site
interactions at the ligand sites can be neglected, since we find that
doubly occupied cluster states are very important in the spin exchange
paths. But note that there is also a dilution effect, as the extended
nature of  the cluster states reduces the impact of the on-site
interactions. Unfortunately, this could not be quantified, since we do not
see any simple way to include electronic interactions at ligand sites
into our perturbative (cluster) treatment. Moreover, the magnitude of
the interaction parameters are much less established than those for
Cu$^{2+}$. Exact diagonalization studies are limited to a few Cu and
O(2) atoms and also have to rely on estimates of the O(2) interaction
parameters. Density functional theory calculations of $J_1$ and $J_2$
are hampered by the problem that the anti-ferromagnetic state is
not the DFT ground state of the solid. In our view, the only way to
improve the present level of quality in magnetic coupling calculations
is to use ab initio quantum chemistry methods
for small $[\mbox{CuO}_2]_n$ clusters, properly embedded.   

In view of the above discussion we think that our method is a reliable
tool for the calculation of magnetic coupling constants and related
quantities such as spin-phonon couplings in relatively complex materials like
CuGeO$_3$. The main advantage is that band-like properties of the
system under consideration are taken into account. In addition, it
allows to investigate the distance dependence of coupling constants in
an easy manner. 

\begin{acknowledgements}
We acknowledge that part of the project has been supported by the
Deutsche Forschungsgemeinschaft.
\end{acknowledgements}
\end{multicols}
%
%
%
%
%
%
\begin{figure}[p]
\begin{center}
\ \epsfxsize= 8 cm \epsffile{jeta.epsi}
\end{center}
\narrowtext
\caption{Dependence of $J_1$ on the Cu-O(2)-Cu angle ${\eta}$ for out of
phase ($q_z = {\pi}$, full line) and in phase ($q_z= 0$, broken
line) distortions of neighboring cells. At the minima around
$84^{\circ}$ ($88^{\circ}$) $J_1 $ is slightly negative, due to
interference effects discussed in
Ref.~\protect\onlinecite{FeWe98}. The vertical line at $\sim
99^{\circ}$ indicates the equilibrium angle. The dot-dashed curve
shows the results of Ref.~\protect\onlinecite{brad98}.}   
\label{jeta}
\end{figure}
%
%
%
%
\begin{figure}[p]
\begin{center}
\ \epsfxsize= 8 cm \epsffile{j1j2p.epsi}
\end{center}
\narrowtext
\caption{Dependence of $J_1$ (upper part) and $J_2$ (lower part) on
pressure. The squares (triangles) show the values of $J_1$ ($J_2$)
calculated from the experimental structure parameters at the
respective pressures. The full line shows $J_i(p)$ calculated from a
linear fit to the experimental crystal data of
Ref.~\protect\onlinecite{brad99}, the dashed lines represent the
$J_i(p)$ of the non-linear fit.}  
\label{jp}
\end{figure}  
%
%
%
%
%
%
\begin{table}
\narrowtext
\caption{Parameter sets A, B, and C for CuGe$_2$O$_3$ in eV and
eV/{\AA} for ${\beta}$ parameters. Subscript 1 represents O(1),
Subscript 2 represents planar O(2).}
\label{cugeoparsets}\vspace{2mm} \centering
\ 
\begin{array}[b]{cccccccccccccc}
\hline
\hline
&U&J_H&{\epsilon}_1&pd{\sigma}_1&pd{\pi}_1
&pd{\sigma}_2&pd{\pi}_2&pp{\sigma}&pp{\pi}
&{\beta}^{pd}_{\sigma}&{\beta}^{pd}_{\pi}
&{\beta}^{pp}_{\sigma}&{\beta}^{pp}_{\pi} \\ 
\hline 
A&8&-0.8&-&  -  & - &-1.2&0.4&-1.0&0.4&1.5&2&2&2.5\\
B&8&-0.8&3&-0.35&0.4&-1.2&0.4&-1.0&0.4&1.5&2&2&2.5\\
C&8&-0.8&3&-0.08&0.4&-1.2&0.4&-0.6&0.4&1.5&2&2&2.5\\
\hline
\hline
\end{array}
\end{table}
%
%
\begin{table}
\narrowtext
\caption{Non-zero Crystal field matrix elements for O(2) p orbitals
(in eV) from the point charge model described in Sec.III.}
\label{penerg}\vspace{2mm} \centering
\ 
\begin{array}[b]{cccc}
\hline
\hline
V^{xx}&V^{yy}&V^{zz}&V^{xy} \\
\hline
      -0.3    &     +0.2     &   +0.1 & -0.2       \\
\hline
\hline
\end{array}
\end{table}        
%
%
\begin{table}
\narrowtext
\caption {Energies of the symmetry adapted Cu d orbitals (in eV). }
\label{denerg}\vspace{2mm} \centering
\ 
\begin{array}[b]{cccccc}
\hline
\hline
 &{\epsilon}_d^1&{\epsilon}_d^2&{\epsilon}_d^3&{\epsilon}_d^4&{\epsilon}_d^5
\\
\hline
A16& -3.5 & -0.4 & -0.1 & -0.1 & -0.1   \\
B16& -3.6 & -0.8 & -0.5 & -0.4 & -0.1   \\
C16& -2.4 & -0.8 & -0.5 & -0.4 & -0.1   \\
\hline
\hline
\end{array}
\end{table}         
%
%
\begin{table}
\narrowtext
\caption{Nearest ($J_1$) and next nearest ($J_2$) neighbor exchange
constants for CuGeO$_3$ (in K) for the sets A, B, and C and for
O(2) cluster sizes of 12 and 16 oxygen atoms, respectively. In
addition, the results using the zero pressure structural parameters of
Ref.~\protect\onlinecite{brad99} are given (B16p). }
\label{js}\vspace{2mm} \centering
\ 
\begin{array}[b]{cccccc}
\hline
\hline
   & A12 & A16 & B16 & C16 & B16p \\
\hline
J_1& 163 & 168 & 159 & 158 & 174  \\
J_2&  53 &  52 &  49 &  48 &  49  \\
\hline
\hline
\end{array}
\end{table}         
%
%
\begin{table}[t]
\narrowtext
\caption{Spin phonon coupling constants for different lattice
distortions (in K/\AA). The superscript labels the
crystallographic directions $a, b, c$, the subscript labels the
respective ion. For comparison, the result of WGB are also shown.}   
\label{coups}\vspace{2mm} \centering
\ 
\begin{array}[b]{cccccc}
\hline
\hline
g^{i}      &  A12     &  A16    &  B16     &  C16    & WGB       \\
\hline
g^c_{Cu}   & -271     & -278    & -274     & -319    & -890      \\
g^b_{Ge}   &   32     &   34    &   32     &   49    & -110      \\
g^a_{O(2)} &  397     &  403    &  393     &  610    &  400      \\
g^b_{O(2)} & -290     & -295    & -280     & -452    & - 91      \\ 
\hline
\hline
\end{array}
\end{table}         
%
%
\begin{table}[t]
\narrowtext
\caption{Coupling constants to angles and distances obtained from a linear
transformation of the $g^i$ (except $g^d_{\eta}$ and $g^d_{dCu}$,
which have been calculated directly from the respective distortion
pattern) in K/\AA. For better comparison, the angular coupling
constants $g_{\eta}$ and $g_{\delta}$
are also given in K/\AA, where we related the change of $J_1$ with
angle to the distance d between the two outer ions defining the angle
via $g_{\phi} = \frac{{\partial}J_1}{{\partial}{\phi}}
\frac{{\partial}{\phi}}{{\partial}d}$.}    
\label{coupsphys}\vspace{2mm} \centering
\ 
\begin{array}[b]{cccccc}
\hline
\hline  
g_{i}      &  A12     &  A16    &  B16     &  C16    & WGB       \\
\hline              
g_{\eta}   &  400     &  407    &  395     &  566    &  708      \\
g_{\eta}^d &          &  413    &  406     &  582    &           \\
g_{\delta} & - 15     & - 16    & -  0.2   & -  5    &  224      \\ 
g_{dCu}    & - 48     & - 48    & - 43     & -114    &  180      \\
g^d_{dCu}  & - 48     & - 47    & - 47     & -116    &           \\ 
g_{dGe}    &   37     &   38    &   39     &   54    & - 96      \\ 
\hline
\hline
\end{array}
\end{table}         
%
%
\begin{table}[t]
\narrowtext
\caption{Coupling constants $g_n, n={1,2,3,4}$
(in K) of the four Peierls-active modes. }  
\label{coupssp}\vspace{2mm} \centering
\ 
\begin{array}[b]{cccccc}
\hline
\hline 
g_{i}      &  A12     &  A16    &  B16     &  C16    & WGB       \\
\hline              
g_1        & -  1.5   & -  1.5  & -  2.2   & -  1.6  & - 15      \\
g_2        &   26     &   26    &   26     &   35.9  &   58      \\
g_3        & -  8     & -  8    & -  8     & -  8.2  & - 30      \\
g_4        & - 22     & - 22    & - 21     & - 34.1  & - 12      \\ 
\hline
\hline
\end{array}
\end{table}                        
%
%
\begin{table}
\narrowtext
\caption {Spin phonon coupling constants of $J_1$ and $J_2$ (in
K/{\AA}) for the $(q_z=0)$ displacements, using set B16.}
\label{coupgz}\vspace{2mm} \centering
\ 
\begin{array}[b]{ccccc}
\hline
\hline
 &g^a_{O(2)}&g^b_{O(2)}&g^{b}_{Ge}&g_{\eta} \\
\hline
J_1 &  258 & -180 & 56 & 460  \\
J_2 & -167 &  119 &  2 &  92  \\
\hline
\hline
\end{array}
\end{table}                 
%
%
\begin{table}
\narrowtext
\caption {Pressure gradients of the seven crystal parameters at p=0
from linear and non-linear fits to the data (in
{\AA}/GPa for $a$, $b$, and $c$ and in 1/GPa for the internal lattice
parameters).} 
\label{prcoup}\vspace{2mm} \centering
\ 
\begin{array}[b]{cccccccc}
\hline
\hline
 &a&b&c&xGe/a&xO1/a&xO2/a&yO2/b
\\
\hline
\mbox{lin}  &-0.013 &-0.12 &-0.0045 &-0.0054 &-0.0053 &-0.0037&-0.0015 \\
\mbox{n-lin}&-0.021 &-0.17 &-0.0044 &-0.0093 &-0.0062 &-0.010 &-0.0024 \\
\hline
\hline
\end{array}
\end{table}         
%
%
%
%
%


\begin{references}
\bibitem{oxhop}
The magnitude of the O-O hopping is supposed to be very similar to the
O-O hopping in the cuprate planes.
\end{references}
\end{document}